\documentclass[a4paper]{jpconf}
\usepackage[breaklinks,colorlinks,bookmarks=false,citecolor=blue,linkcolor=red,urlcolor=blue]{hyperref}
\usepackage{iopams}
\usepackage{graphicx}
\usepackage{cite}
\usepackage{wrapfig}
\def\abs#1{\left|#1\right|}

\begin{document}
\title{Quantum Monte Carlo simulations of highly frustrated magnets in a 
cluster basis: The two-dimensional Shastry-Sutherland model}

\author{Andreas Honecker$^1$, Lukas Weber$^2$, Philippe Corboz$^3$,
 Fr\'ed\'eric Mila$^4$, Stefan Wessel$^2$}

\address{$^1$ Laboratoire de Physique Th\'eorique et Mod\'elisation, CNRS UMR 8089, CY Cergy Paris Universit\'e, Cergy-Pontoise, France}
\address{$^2$ Institute for Theoretical Solid State Physics, RWTH Aachen University, Germany}
\address{$^3$ Institute for Theoretical Physics, University of Amsterdam, The Netherlands}
\address{$^4$ Institute of Physics, Ecole Polytechnique F\'ed\'erale de Lausanne (EPFL), Switzerland}

\ead{andreas.honecker@cyu.fr}

\begin{abstract}
Quantum Monte Carlo (QMC) simulations constitute nowadays one of the most 
powerful methods to study strongly correlated quantum systems, provided 
that no ``sign problem'' arises. However, many systems of interest, 
including highly frustrated magnets, suffer from an average sign that is 
close to zero in standard QMC simulations. Nevertheless, a possible sign 
problem depends on the simulation basis, and here we demonstrate how a 
suitable choice of cluster basis can be used to eliminate or at least 
reduce the sign problem in highly frustrated magnets that were so far 
inaccessible to efficient QMC simulations. We focus in particular on the 
application of a two-spin (dimer)-based QMC method to the thermodynamics 
of the spin-1/2 Shastry-Sutherland model for SrCu$_2$(BO$_3$)$_2$.
\end{abstract}

\section{Introduction}

Quantum many-body systems constitute a challenge for a quantitatively 
accurate treatment. In this context, Quantum Monte Carlo (QMC) methods 
have become one of the most powerful tools 
\cite{Evertz03,Troyer03,Sandvik10}. However, in some cases, including 
geometrically frustrated magnets, their use is limited by a phenomenon 
known as the ``sign problem'', namely a cancellation of positive and 
negative weights of configurations \cite{Troyer05,Hangleiter20,Hen21}. In 
the present contribution, we illustrate how these problems can be overcome 
or at least alleviated by working in a suitable cluster basis 
\cite{Nakamura98,Honecker16,Alet16,Weber21}. For the purpose of 
simplicity, we will focus on the application of a dimer basis to a highly 
frustrated two-dimensional spin-1/2 model known as the 
``Shastry-Sutherland model'' \cite{Shastry81} that is realised in the 
compound SrCu$_2$(BO$_3$)$_2$ \cite{Kageyama99,Miyahara03}.

\section{Method and the Shastry-Sutherland model}

\label{sec:Method}

One flexible QMC approach is based on the representation of the partition 
function $Z$ as a high-temperature series \cite{Handscomb64,Sandvik92}, 
leading to the ``stochastic series expansion'' (SSE). The SSE is based on 
a splitting of the Hamiltonian $H$ into bond terms $H_b$. These terms are 
split in turn into different types, such that the action of each bond term 
$H_{(b,t)}$ on a given basis state $| \alpha \rangle$ either vanishes or 
yields exactly one other basis state $| \alpha' \rangle$. Note that one of 
the types $t$ is diagonal while all others are off-diagonal. Using $H = 
\sum_{(b,t)} H_{(b,t)}$, one finds in the given basis
\begin{equation}
Z = {\rm Tr} \, {\rm e}^{-\beta\,H}
  = \sum_{n=0}^{\infty} \frac{\beta^n}{n!} \, 
     \sum_{| \alpha_n \rangle = | \alpha_1 \rangle}
     \sum_{(b_1,t_1), \ldots, (b_n,t_n)}
     \prod_{i=1}^n \langle \alpha_{i+1} | \left(-H_{(b_i,t_i)}\right) | \alpha_{i} \rangle
  = \sum_c W_c
\, ,
\label{eq:SSEZ}
\end{equation}
where $\beta = 1/T$ is the inverse temperature (we set the Boltzmann 
constant $k_B = 1$). Important developments concern the representation and 
efficient Monte Carlo sampling of the sum in Eq.~(\ref{eq:SSEZ}), as well 
as measurement of physical quantities, and we refer to 
Refs.~\cite{Syljuasen02,Syljuasen03,Alet05,Weber21} for detailed 
discussions of these issues.

Here, our main concern will be a different one: the weights $W_c$ of a 
configuration $c$ in Eq.~(\ref{eq:SSEZ}) are given by products of matrix 
elements $\langle \alpha_{i+1} | \left(-H_{(b_i,t_i)}\right) | \alpha_{i} 
\rangle$ and there is in general no guarantee that they are non-negative, 
{\it i.e.}, that the expression Eq.~(\ref{eq:SSEZ}) does indeed have a 
probabilistic interpretation. For diagonal terms of the Hamiltonian, one 
can add a constant shift $\epsilon$ to render $-H_{(b_i,t_i)} + \epsilon$ 
non-negative. When off-diagonal terms are negative, one uses the absolute 
value of the weights $\abs{W_c}$ for sampling such that the average of an 
observable $A$ is given as the ratio of the averages of the observable and 
of the sign \cite{Evertz03,Troyer05},
\begin{equation}
\langle A \rangle = \frac{\sum_c W_c \, A_c}{\sum_c W_c} = \frac{\sum_c 
{\rm sign}(W_c) \, \abs{W_c} \,A_c}{\sum_c {\rm sign}(W_c)\,\abs{W_c}} = \frac{\langle 
{\rm sign} \, A \rangle_{\abs{\cdot}}}{\langle {\rm sign} \rangle_{\abs{\cdot}}} \, .
\label{eq:sign}
\end{equation}
This allows one to run a QMC simulation for any given problem. However, if 
indeed negative contributions appear in the sum in Eq.~(\ref{eq:SSEZ}), 
there is a cancellation of negative and positive contributions to $\langle 
{\rm sign} \rangle_{\abs{\cdot}}$, and if these are comparable in 
magnitude one may not be able to distinguish the normalisation $\langle 
{\rm sign} \rangle_{\abs{\cdot}}$ from zero within the error bars of the 
calculation, thus rendering the results meaningless. This problem is known 
as the ``sign problem'', and we will provide some examples in section 
\ref{sec:Results}. Nevertheless, because the sign of the terms in 
Eq.~(\ref{eq:SSEZ}) depends on the choice of basis, one may be able to 
eliminate or at least alleviate it by a simulation basis that is suitably 
chosen for a given problem, and we will illustrate one such approach in 
the following.

\begin{wrapfigure}{r}{0.4\textwidth}
\vspace*{-4.5mm}
 \centering
  \includegraphics[width=0.3\textwidth]{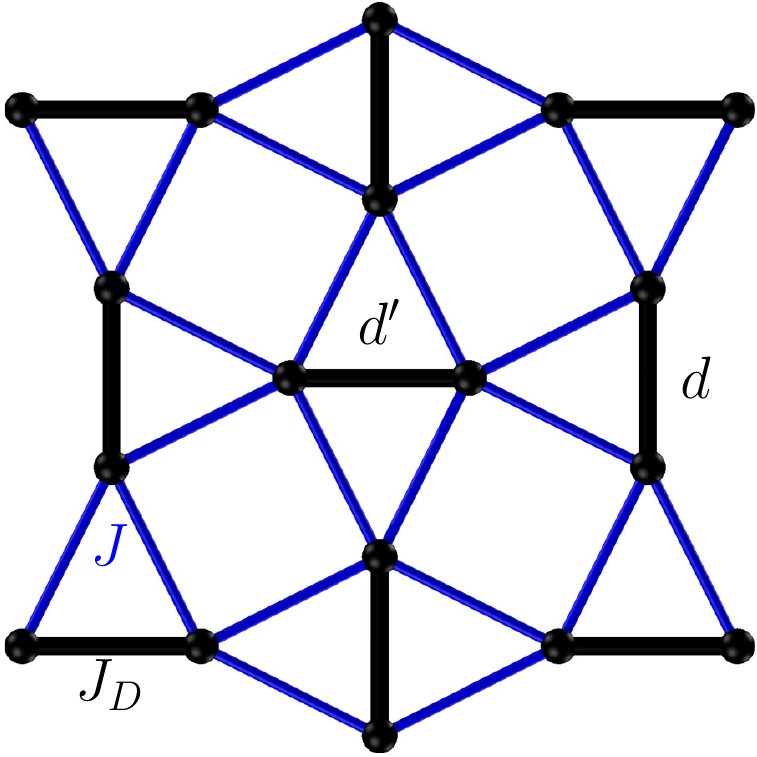}
\vspace*{-1.5mm}
  \caption{Schematic representation of the Shastry-Sutherland model. The 
  black dots correspond to sites carrying spins 1/2. Dimer couplings $J_D$ 
  are shown by thick black lines and inter-dimer couplings $J$ by thinner 
  blue lines. Two neighbouring dimers are designated by $d$ and $d'$.
 \label{fig:ShaSu}}
\vspace*{-9mm}
\end{wrapfigure}
In order to be specific, we will now focus on the Shastry-Sutherland 
model, which is represented schematically in Fig.~\ref{fig:ShaSu}. The 
Hamiltonian in the single-site basis is
\begin{equation}
H = J_D \, \sum_{\langle i,j \rangle} \vec S_i \cdot \vec S_j + J \, \sum_{\langle
\langle i,j \rangle\rangle} \vec S_i \cdot \vec S_j \, ,
\label{eq:HShaSu}
\end{equation}
where $\vec S_i$ are spin-1/2 operators for site $i$. $J_D$ is the 
intra-dimer coupling (denoted by $\langle i,j \rangle$) and the 
inter-dimer coupling ($\langle\langle i,j \rangle\rangle$) $J$ corresponds 
to a square lattice, as shown in Fig.~\ref{fig:ShaSu}. This model was 
constructed \cite{Shastry81} to ensure that the ground state is an exact 
product of singlets formed on the dimer bonds for small and intermediate 
coupling ratios $J/J_D$. More recently, this model was also found to be 
realised in the quantum spin system SrCu$_2$(BO$_3$)$_2$ 
\cite{Kageyama99,Miyahara03}, which increased the interest in calculating 
its thermodynamic properties.

As a consequence of the triangular arrangement of the $J$ and $J_D$ 
interactions, this problem is geometrically frustrated. Consequently, QMC 
simulations in the single-spin basis (\ref{eq:HShaSu}) suffer from a 
severe sign problem, as we will show in section \ref{sec:Results}. 
However, one may pass to the total-spin and spin-difference operators, 
respectively $\vec T_d = \vec S_{d,1} + \vec S_{d,2}$ and $\vec D_d = \vec 
S_{d,1} - \vec S_{d,2}$, for the two spins $\vec S_{d,1}$ and $\vec 
S_{d,2}$ on a given dimer $d$ \cite{Wessel18}. The coupling of the two 
spins in dimer $d$ can then be expressed through $\vec T_d^2$ and the 
inter-dimer coupling for the two dimers $d$ and $d'$ indicated in 
Fig.~\ref{fig:ShaSu} takes the form
\begin{equation}
H_{dd'} = {\textstyle \frac12}\,J\,
\vec T_d \cdot \vec T_{d'}
 - 
 {\textstyle \frac12}\,J\,
\vec T_d \cdot \vec D_{d'}
\, .
\label{eq:Hdd}
\end{equation}
The first term corresponds just to a composite-spin model on a square 
lattice, which is unfrustrated and thus poses no QMC sign problem. The 
sign of the coefficient of the second term in Eq.~(\ref{eq:Hdd}) depends 
on the assignment of the labels $1$ and $2$ in dimer $d'$ and varies for 
the different inter-dimer coupling terms throughout the lattice. There are 
also signs involved in the construction of the operator $\vec D_{d’}$ such 
that it is not immediately clear to which extent this second term may 
still give rise to a sign problem. Nevertheless, the operator $\vec D_d$ 
yields dimer triplet states when applied to the singlet state of the dimer 
$d$, $|S \rangle_{d} = \frac{1}{\sqrt{2}} \, \left( |\uparrow \downarrow 
\rangle_{d} - |\downarrow \uparrow \rangle_{d} \right)$. Because the 
ground state is an exact product of such dimer singlets for $J \ll J_D$, 
one may expect the difference operators $\vec D_d$ not to contribute to 
the leading orders at low temperatures, such that a potential sign problem 
remains manageable. We will show in the next section that indeed this 
turns out to be true in the regime where $J$ is not too large.

\section{Results}

\label{sec:Results}

We now present some QMC results for the Shastry-Sutherland model where we 
will focus on the two representative cases $J/J_D = 0.5$ and $0.55$.

\begin{figure}[t!]
\centering
\includegraphics[width=0.48\columnwidth]{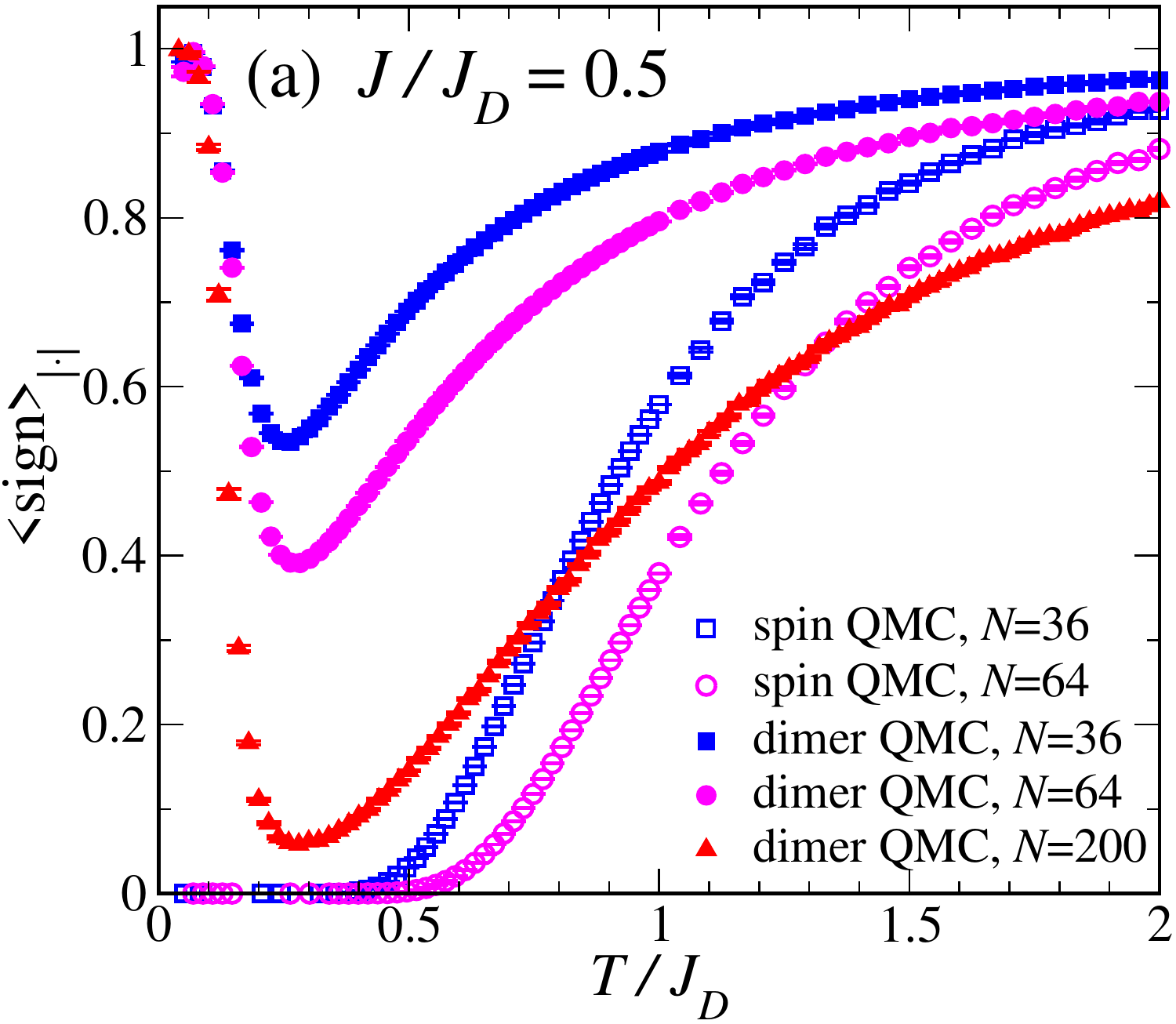}%
\hfill%
\includegraphics[width=0.48\columnwidth]{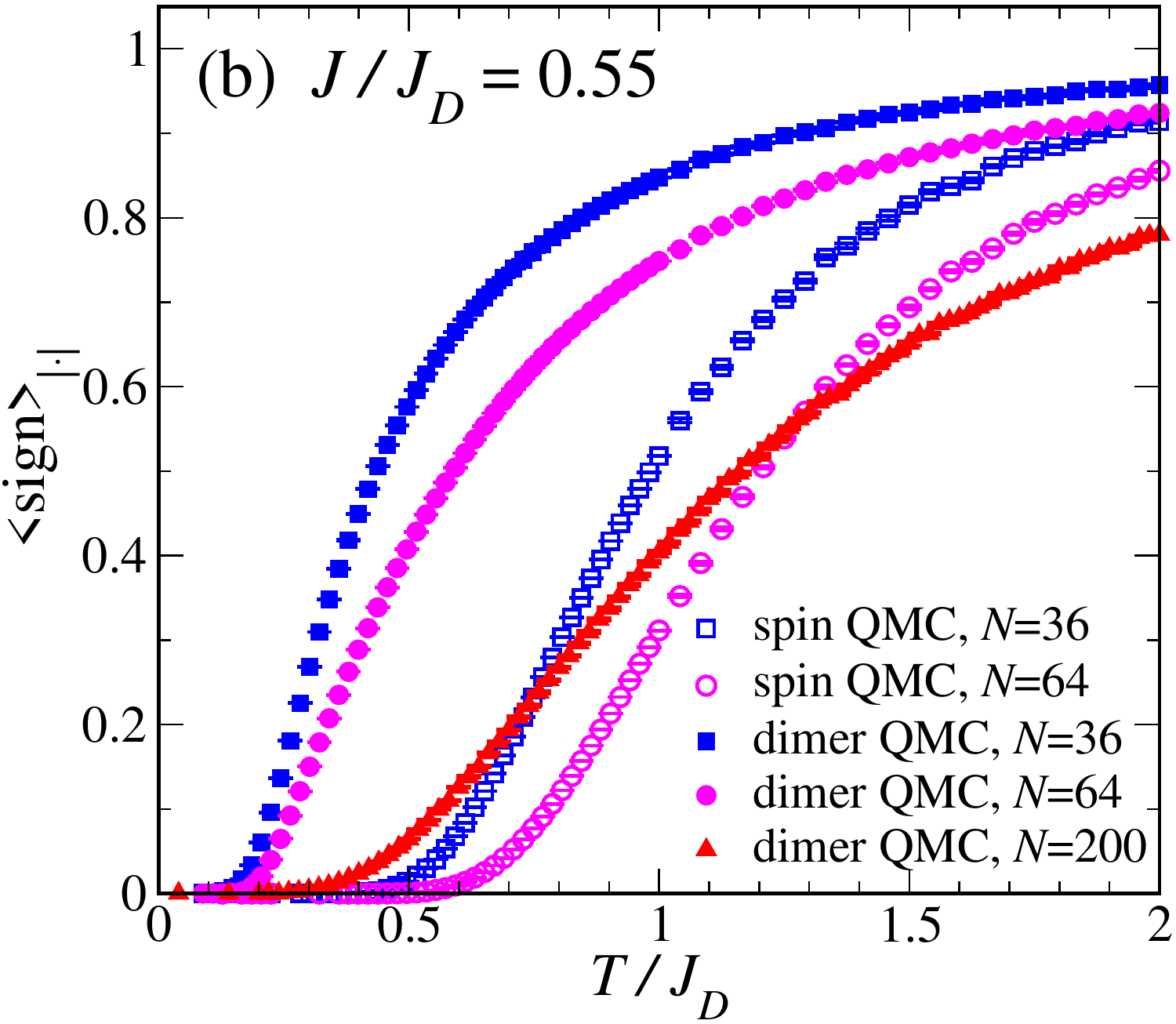}
\vspace*{-2.5mm}
\caption{Average sign $\langle {\rm sign} \rangle_{\abs{\cdot}}$ computed 
as a function of temperature for the Shastry-Sutherland model with (a) 
$J/J_D = 0.5$ and (b) $J/J_D = 0.55$. Open symbols are for the single-spin 
basis, filled symbols for the dimer basis. Data for $N=200$ spins are 
reproduced from Ref.~\cite{Wessel18}.
\label{fig:sign}}
\vspace*{-2.5mm}
\end{figure}

Figure~\ref{fig:sign} presents the average sign $\langle {\rm sign} 
\rangle_{\abs{\cdot}}$. In the single-spin basis (open symbols), the 
geometric frustration leads to the expected sign problem (see section 
\ref{sec:Method}), which manifests itself as the rapid drop of the average 
sign at low temperatures. If a sign problem arises, the average sign 
generically becomes exponentially small in $N/T$ \cite{Evertz03,Troyer05}. 
This is verified by the single-site results, where the average sign not 
only drops at low temperatures but also decreases with increasing system 
size $N$. Working in the dimer basis improves the sign problem very 
significantly, as is illustrated by the full symbols in 
Fig.~\ref{fig:sign}. There is still a sign problem, as evidenced by the 
regime in which $\langle {\rm sign} \rangle_{\abs{\cdot}} < 1$. However, 
for $N=32$ and $64$ spins, the average sign is always bigger in the dimer 
basis than in the single-site basis. Most remarkably, at least in the case 
$J/J_D = 0.5$ (Fig.~\ref{fig:sign}(a)) the average sign does not vanish as 
temperature goes to zero, but instead recovers after going through a 
minimum around $T/J_D \approx 0.25$. This minimum still drops with 
increasing $N$, but one can go to systems as large as $N=200$ and still 
ensure that $\langle {\rm sign} \rangle_{\abs{\cdot}} \gtrsim 0.06$, such 
that the resulting loss of accuracy can be compensated by improving the 
statistics.

At $J/J_D = 0.55$ (Fig.~\ref{fig:sign}(b)), the average sign no longer 
recovers for $T \to 0$, even in the dimer basis. This difference between 
the cases $J/J_D = 0.5$ and $0.55$ can be traced to an algorithmic phase 
transition at $J/J_D = 0.526(1)$ \cite{Wessel18}, where the effective 
model defined by the absolute value of the weights in Eq.~(\ref{eq:sign}) 
undergoes a transition out of the dimer phase. Nevertheless, $\langle {\rm 
sign} \rangle_{\abs{\cdot}}$ in the dimer basis for $N=200$ spins retains 
a larger value than that in the single-site basis for the much smaller 
system $N=32$ at temperatures $T/J_D \lesssim 0.7$, as shown in 
Fig.~\ref{fig:sign}(b).

\begin{figure}[t!]
\centering
\includegraphics[width=0.48\columnwidth]{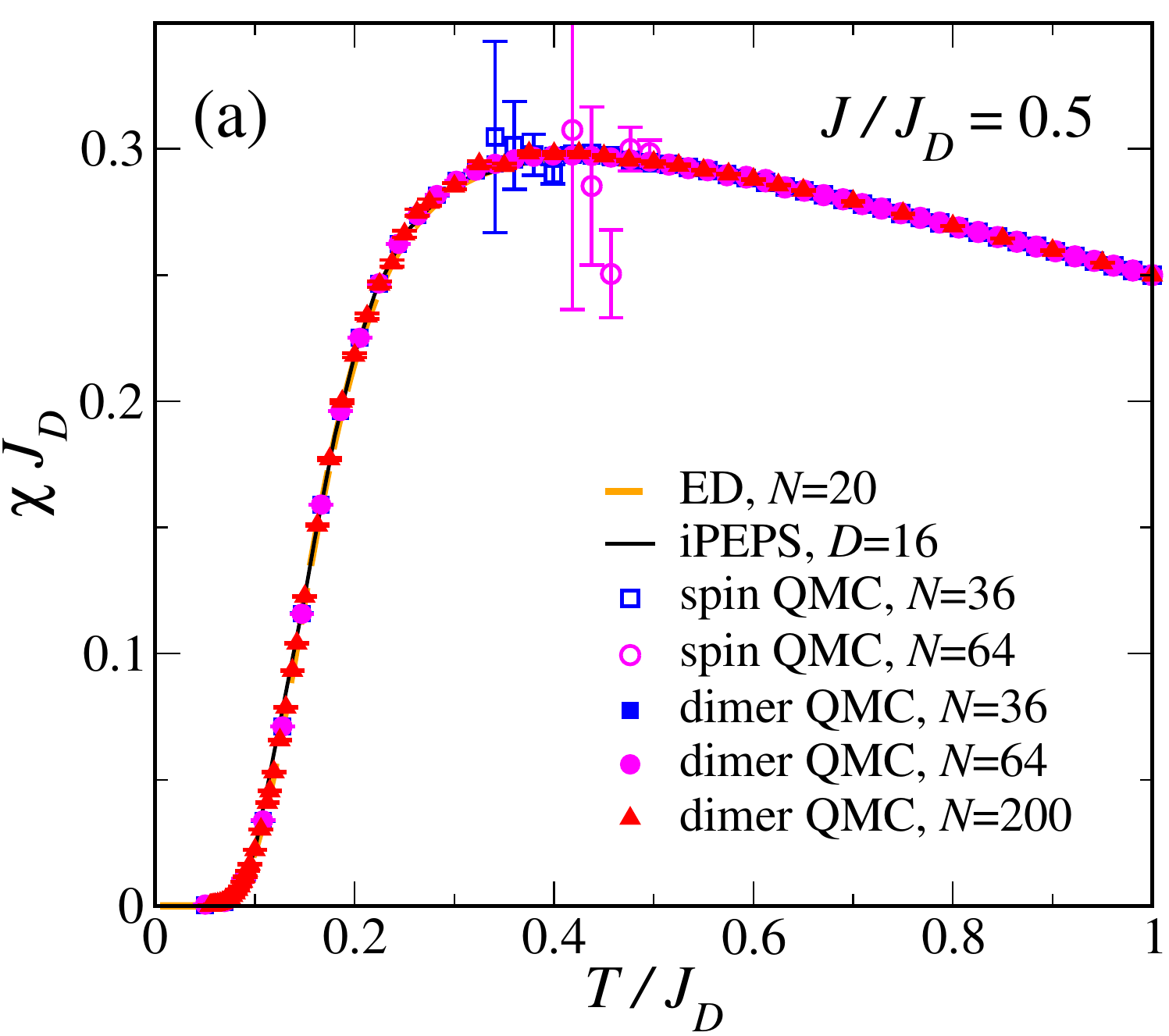}%
\hfill%
\includegraphics[width=0.48\columnwidth]{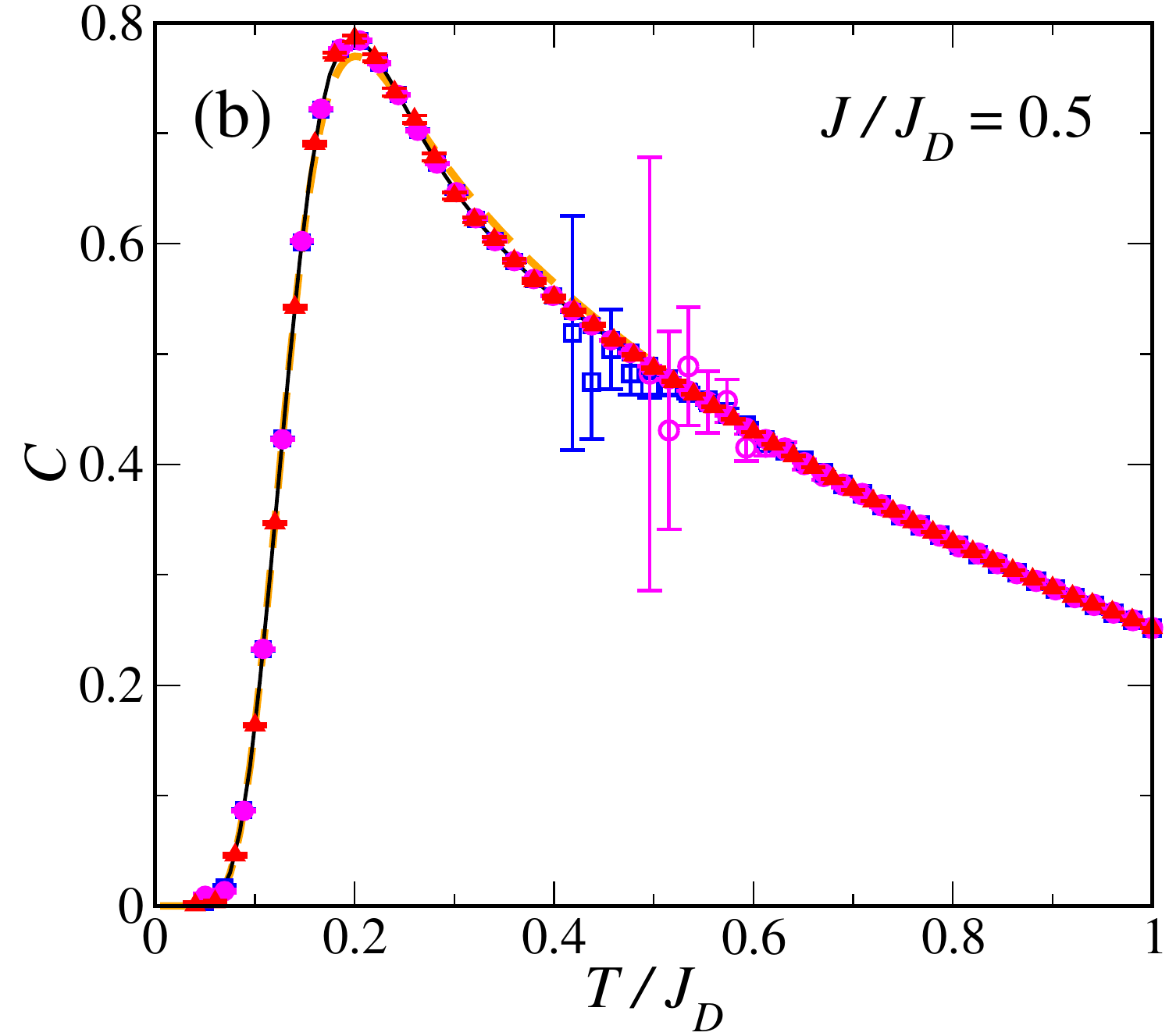}
\vspace*{-2.5mm}
\caption{(a) Magnetic susceptibility $\chi$ and (b) specific heat $C$ per 
dimer of the Shastry-Sutherland model at $J/J_D = 0.5$. Exact 
diagonalisation (ED) results for $N=20$ and QMC data for $N=200$ spins are 
reproduced from Ref.~\cite{Wessel18}, $D=16$ iPEPS data from 
Ref.~\cite{Wietek19}.
\label{fig:chiC0_5}}
\vspace*{-2.5mm}
\end{figure}

\begin{figure}[t!]
\centering
\includegraphics[width=0.48\columnwidth]{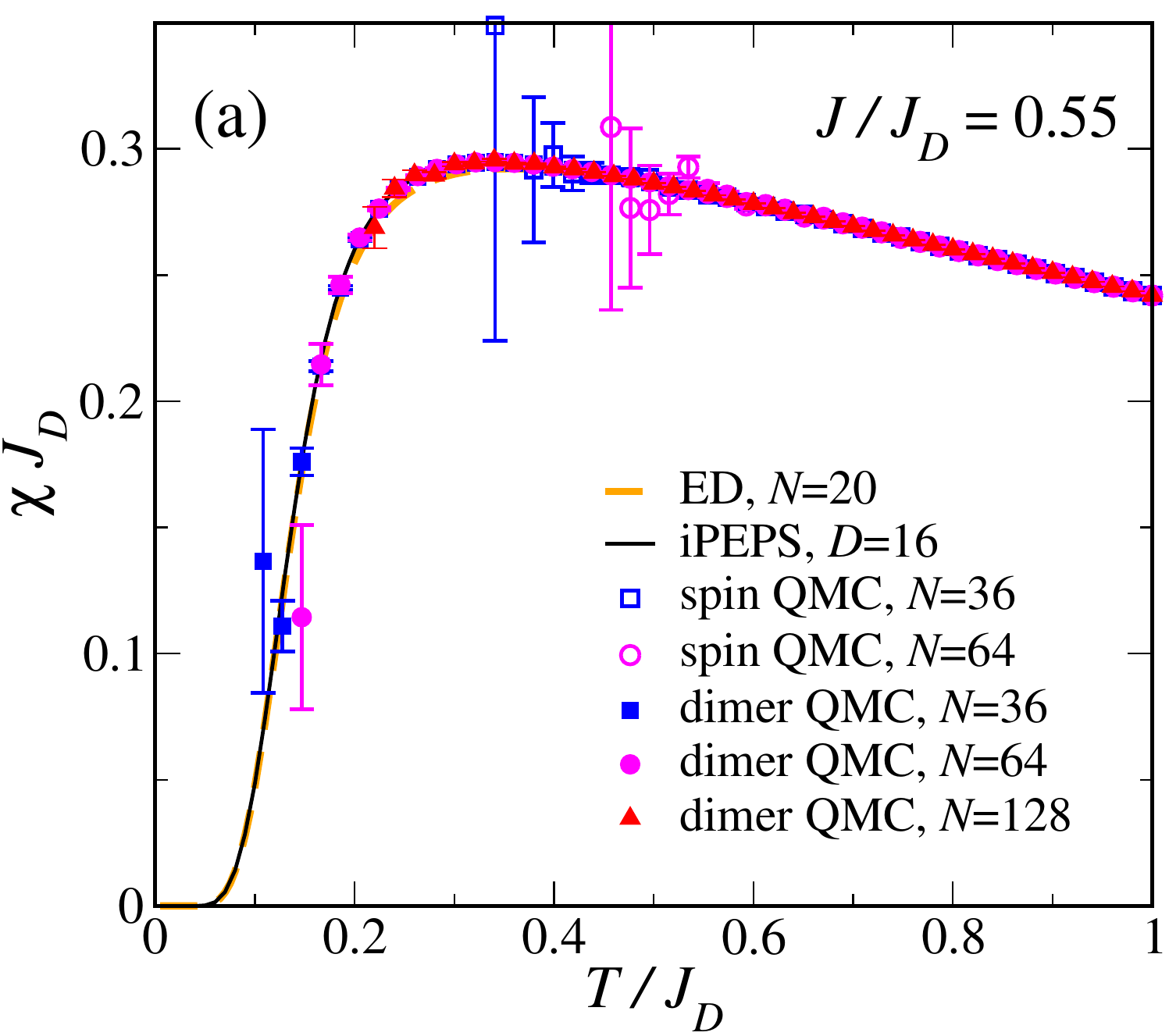}%
\hfill%
\includegraphics[width=0.48\columnwidth]{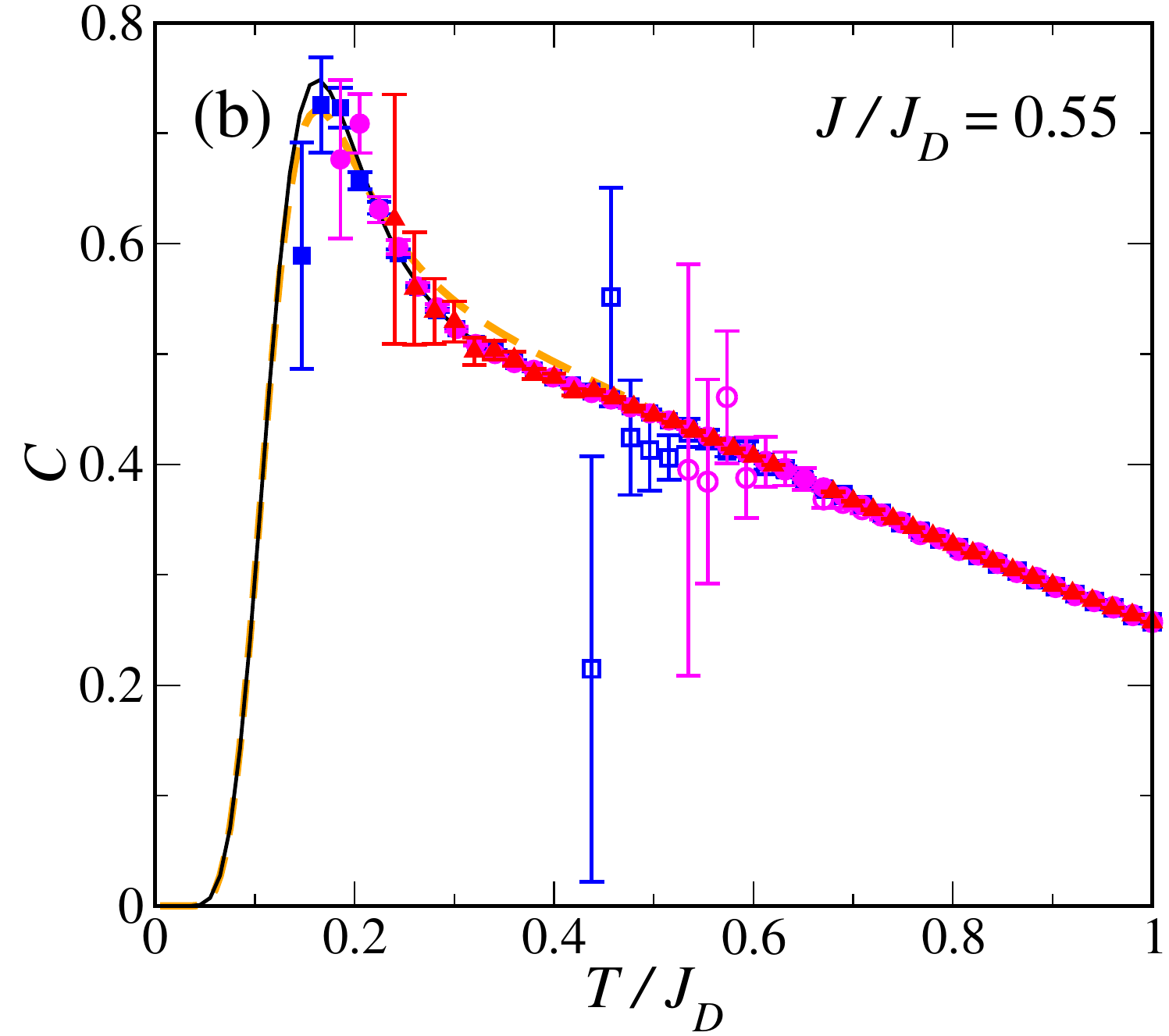}
\vspace*{-2.5mm}
\caption{Same as Fig.~\ref{fig:chiC0_5}, but for $J/J_D = 0.55$. Exact 
diagonalisation (ED) results for $N=20$ and QMC data for $N=128$ spins are 
reproduced from Ref.~\cite{Wessel18}.
\label{fig:chiC0_55}}
\vspace*{-2.5mm}
\end{figure}

Next we present results for thermodynamic observables at $J/J_D = 0.5$ and 
$0.55$ which are shown respectively in Figs.~\ref{fig:chiC0_5} and 
\ref{fig:chiC0_55}. As before, QMC results in the single-spin basis are 
shown by open symbols and results obtained in the dimer basis by filled 
symbols. For reference, we have included exact diagonalisation (ED) 
results for $N=20$ spins \cite{Wessel18} and infinite projected 
entangled-pair states (iPEPS) data for a bond dimension $D=16$ 
\cite{Wietek19}. As thermodynamic observables we have selected the 
magnetic susceptibility $\chi$ and specific heat $C$; the temperature 
window $0 \le T \le J_D$ is chosen to include the maximum of these two 
quantities.

The single-site basis is barely able to resolve the maximum of $\chi$ at 
$J/J_D=0.5$ (Fig.~\ref{fig:chiC0_5}(a)) or get close to it for 
$J/J_D=0.55$ (Fig.~\ref{fig:chiC0_55}(a)), provided that one restricts a 
system size as small as $N=36$. However, single-site QMC generally loses 
accuracy already at significantly higher temperatures, such that the 
maxima are out of reach and only the high-temperature regime is 
accessible, as was to be expected according to the average sign shown in 
Fig.~\ref{fig:sign}. On the other hand, for $J/J_D = 0.5$, the dimer basis 
allows one to obtain accurate QMC data for systems with up to $N=200$ 
spins and at all temperatures, as shown in Fig.~\ref{fig:chiC0_5}. For 
$J/J_D = 0.55$, the sign problem of Fig.~\ref{fig:sign}(b) prevents one 
from reaching the zero-temperature limit, but the maximum of the magnetic 
susceptibility can still be well resolved in the dimer basis for $N\le 
128$ spins (Fig.~\ref{fig:chiC0_55}(a)). In order to resolve the maximum 
of $C$ (Fig.~\ref{fig:chiC0_55}(b)) by QMC in the dimer basis, one needs 
to compromise on system size, {\it i.e.}, restrict to $N=36$, and on 
accuracy, but even $N=128$ still works significantly better in the dimer 
basis than $N=36$ in the single-spin basis, again as expected from the 
results for the average sign shown in Fig.~\ref{fig:sign}.

The iPEPS data for $D=16$ can be considered as representative of the 
thermodynamic limit in the present parameter regime and at the level of 
resolution in Figs.~\ref{fig:chiC0_5} and \ref{fig:chiC0_55} 
\cite{Wietek19}. Indeed, all of our QMC results agree with iPEPS within 
the corresponding statistical error bars. On the other hand, small (albeit 
distinct) finite-size effects can be seen in the $N=20$ ED data, in 
particular around the maximum of $C$ and at temperatures above it. These 
finite-size effects become increasingly pronounced at the larger value 
$J/J_D = 0.55$ (Fig.~\ref{fig:chiC0_55}(b)). This highlights the 
importance of having access to larger systems in order to ensure 
quantitatively accurate results.

Concerning the physics of these thermodynamic quantities, we observe the 
emergence of low-temperature maxima in the magnetic susceptibility and the 
specific heat that move to lower temperature and sharpen as $J/J_D$ 
increases \cite{Wessel18,Wietek19}. In the present results, this is seen 
most clearly for the specific heat, shown in Figs.~\ref{fig:chiC0_5}(b) 
and \ref{fig:chiC0_55}(b).

\section{Conclusion}

We have shown how a dimer basis can be used to perform accurate QMC 
simulations for the thermodynamic properties of a highly frustrated 
two-dimensional spin-1/2 model, the Shastry-Sutherland model, that would 
be out of reach of conventional approaches. The compound 
SrCu$_2$(BO$_3$)$_2$ is believed to correspond to the parameter ratio 
$J/J_D \approx 0.63$ \cite{Miyahara03,Wietek19}. This ratio is still in 
the dimer phase of the Shastry-Sutherland model, but rather far beyond the 
algorithmic phase transition at $J/J_D = 0.526(1)$ \cite{Wessel18}. Thus 
the parameter regime relevant to SrCu$_2$(BO$_3$)$_2$ remains 
unfortunately inaccessible to accurate QMC simulations even in the dimer 
basis, and one needs to resort to other methods such as iPEPS 
\cite{Wietek19,Jimenez2021}.

Still, there are other cases to which the ideas illustrated here can be 
applied. For example, in the case of a frustrated ladder, only boundary 
terms contribute to the sign problem, such that it remains sufficiently 
mild to allow for a detailed study of the entire phase diagram 
\cite{Honecker16,Wessel17}. Another case is a fully frustrated bilayer 
model, where the counterpart of the $\vec T_d \cdot \vec D_{d'}$ term in 
Eq.~(\ref{eq:Hdd}) is absent and thus the dimer basis eliminates the sign 
problem completely \cite{Alet16,Ng17,Stapmanns18}. A detailed QMC study of 
the phase diagram then becomes possible, allowing one to trace a 
first-order phase transition from zero temperature to finite temperature 
until it terminates at an Ising critical point 
\cite{Stapmanns18,Jimenez2021}.

The cluster-basis approach is is not restricted to dimers
and we have demonstrated recently that it also works with a trimer 
(three-spin) basis \cite{Weber21}. This permits one to generalise the 
investigation of the fully frustrated bilayer model to a trilayer 
analogue, and study for example the influence of the ground-state entropy 
that arises in the latter model on the slope of the first-order line at 
finite temperatures \cite{Weber21}.

\ack
We would like to thank Bruce Normand for numerous collaborations on 
related problems and for a critical reading of the manuscript.

\section*{References}

\bibliography{dimerQMCproc}

\end{document}